\documentstyle[12pt,epsfig]{article}
\newcommand\MCFM{\tt MCFM}
\def\cA{{\cal A}}
\def\cC{{\cal C}}
\def\cP{{\cal P}}
\def\tree{{\rm tree}}
\def\qb{{\bar q}}
\def\ub{{\bar u}}
\def\db{{\bar d}}
\def\nub{{\bar \nu}}
\def\ellb{{\bar \ell}}
\def\sstw{\sin^2\theta_W}
\def\spa#1.#2{\left\langle#1 \hskip .15 mm #2\right\rangle}
\def\spb#1.#2{\left[#1 \hskip .15 mm #2\right]}
\def\spab#1.#2.#3{\langle\mskip-1mu{#1} 
                  | #2 | {#3}\mskip-1mu\rangle}
\def\gev{\rm GeV}

\newcommand{ \slashchar }[1]{\setbox0=\hbox{$#1$}   
   \dimen0=\wd0                                     
   \setbox1=\hbox{/} \dimen1=\wd1                   
   \ifdim\dimen0>\dimen1                            
      \rlap{\hbox to \dimen0{\hfil/\hfil}}          
      #1                                            
   \else                                            
      \rlap{\hbox to \dimen1{\hfil$#1$\hfil}}       
      /                                             
   \fi}                                             %

\def\etmiss{\slashchar{E}_{T}}

\begin{document}

\begin{titlepage}
\begin{flushright}
FERMILAB--Pub--99/146--T\\
hep-ph/9905386\\
May 1999
\end{flushright}
\begin{center}
\vspace*{2.5cm}
{\Large\bf An update on vector boson pair production \\
at hadron colliders} 
\vskip 0.5cm
{\large J.M. Campbell and R.K. Ellis} \\
\vskip 0.2cm
{Theory Department, Fermilab, PO Box 500, Batavia, IL 60510, USA.}
 
\vskip 4cm
\end{center}
 
\begin{abstract}
\noindent
We present numerical results (including full one-loop QCD corrections) for
the processes $p\bar{p}$ and $pp \rightarrow W^+ W^-, \ W^\pm
Z/\gamma^*$ and $\ Z/\gamma^* \, Z/\gamma^*$ followed by the 
decay of the massive vector bosons into leptons. In
addition to their intrinsic importance as tests of the standard model,
these processes are also backgrounds to conjectured non-standard model
processes. Because of the small cross sections at the Tevatron, full
experimental control of these backgrounds will be hard to achieve.
This accentuates the need for up-to-date theoretical information.
A comparison is made with earlier work and cross section results
are presented for $p \bar{p}$ collisions at $\sqrt{s}=2$~TeV
and $pp$ collisions at $\sqrt{s}=14$~TeV. Practical examples of the 
use of our calculations are presented.
\end{abstract}

\vskip 1cm
\begin{center}
{\sl Submitted to Physical Review D}
\end{center}
 
\end{titlepage}
\setcounter{footnote}{0}
\section{Introduction}
We present results for the hadronic production of
a vector boson pair, including all spin correlations in the 
decay of the final state bosons, 
$q{\bar q} \to V_1V_2 \to {\rm leptons}$ where $V_i = W^\pm, Z$ or 
$\gamma^*$. The calculations are performed in next-to-leading 
order in $\alpha_S$.
We implement the helicity amplitudes of~\cite{dks} 
and thus extend previous treatments of vector boson pair
production (\cite{frixzz}--\cite{frixww} and \cite{ohnezz}--\cite{ohnespin})
to include spin correlations in all the partonic matrix elements. By
including the decay products in this way it is possible to 
impose experimental cuts, necessary to compare theory with experiment. 
Some cuts are experimentally necessary and more stringent cuts are
often useful in order to reduce backgrounds in the 
search for new physics. Although 
phenomenological predictions including the {\it complete} one loop predictions
are presented here for the first time, this may be a matter of theoretical
correctness rather than practical importance. The early predictions 
were performed at $16~{\rm and}~40~{\rm TeV}$ and to a limited extent at 
$1.8~{\rm TeV}$, so an update of the phenomenological results is in 
any case appropriate.
We have therefore provided predictions for $p \bar{p}$
collisions at $2~{\rm TeV}$ (Tevatron Run II) and for $pp$ collisions 
at $14~{\rm TeV}$~(LHC).
Since the early predictions were made, there have been changes in $\alpha_s$
and the determination of the gluon distribution, especially at small $x$.
We include this information by using modern parton distributions.

Our results are obtained using a Monte Carlo program $\MCFM$ 
which allows the calculation of any infra-red finite quantity 
through order $\alpha_s$. 
The Monte Carlo program is constructed using the method of 
Ref.~\cite{CatSey} based on the subtraction technique of Ref.~\cite{ERT}.
We hope to provide further details in a subsequent publication.
 
\section{Total cross sections}
As already noted, there is a substantial existing literature
on vector boson pair production in hadronic collisions. 
As a cross-check of our results, 
we will compare the values of the total cross section 
obtained using our Monte Carlo ($\MCFM$) 
with those of Frixione, Nason, Mele and Ridolfi~\cite{frixzz,frixwz,frixww}.
Since the earlier predictions were made at centre-of-mass energies of
$\sqrt{s}=16$~TeV and $\sqrt{s}=40$~TeV, we will also provide up-to-date values
for $p{\bar p}$ collisions at $2$~TeV and $pp$ collisions at $14$~TeV. These
Run II Tevatron and LHC predictions contain the latest parton distribution
sets~\cite{mrs98,cteq5} as well as more recent electroweak input.

\subsection{Comparison with existing results}
We will first present a comparison with older results on the total 
cross section.
We use the structure function set HMRSB~\cite{hmrsb}, which is common
amongst the calculations of~\cite{frixzz,frixwz,frixww}. 
This set corresponds to a four-flavour value of 
$\Lambda^{\rm QCD}_{(4)} = 190$~MeV.
For the purposes of comparison we use the standard two loop 
expression~\cite{pdg98}
\begin{equation}
\alpha_S(Q^2) = \frac{1} {b \ln ( Q^2 /\Lambda^2)} 
\Bigg[1- \frac{b^\prime}{b}
\frac {\ln \ln ( Q^2 /\Lambda^2)} {\ln ( Q^2 /\Lambda^2)}\Bigg],
\end{equation}
and match to five flavors at $m_b=5~{\rm GeV},
\alpha_s^{(4)}(m_b)=\alpha_s^{(5)}(m_b)$.
This yields a strong coupling at the $Z$-mass of,
\begin{equation}
\alpha_s(m_Z) = 0.10796.
\end{equation}
The renormalization and 
factorization scales are chosen to be equal to the average mass of the
produced boson pair (namely, $m_W$, $m_Z$ and $(m_W+m_Z)/2$).

Finally, since the early calculations present results 
for the production of two
on-shell vector bosons, we need to ensure that in our Monte Carlo (in which
we produce 4 final-state leptons) we use the narrow-width 
approximation where only doubly-resonant diagrams are included. 
The on-shell boson cross sections can be obtained by
dividing out by the relevant branching ratios. 
A full discussion of the other diagrams included in our 
approach is given in Section~\ref{beyondzw}.

The comparison of the results for a $pp$ collider at 
centre of mass energies of 
$\sqrt{s}=16$~TeV and $\sqrt{s}=40$~TeV is shown in Tables~\ref{comp16}
and~\ref{comp40} respectively. For the cases of $WW$ and $ZZ$ pair production,
the values for comparison are taken directly from~\cite{frixzz}
and~\cite{frixww}. For $Z W^\pm $, the results 
are slightly different from those
published in~\cite{frixwz} since the contribution 
from processes of the type $g+b \rightarrow Z +W^-  +t$ 
with the top quark taken massless 
(which it is no longer appropriate to include)
have been 
removed\footnote{We are grateful to P. Nason for providing us with a modified
code for these cases.}. The comparison between our results and the results
of refs.~\cite{frixzz}--\cite{frixww} is satisfactory. Apart from their role 
as a check of our programs the results in Tables~\ref{comp16} and \ref{comp40}
should be considered obsolete.
\begin{table}
\renewcommand\arraystretch{1.1}
\begin{center}
\begin{tabular}{|c||c|c||c|c||c|c||c|c|}
\hline
$\sqrt{s}=16$~TeV & \multicolumn{2}{c||}{$W^+W^-$} & \multicolumn{2}{c||}{$ZW^+$} 
& \multicolumn{2}{c||}{$ZW^-$} & \multicolumn{2}{c|}{$ZZ$} \\
\cline{2-9}
($pp$)& \MCFM & Ref. \cite{frixww} & \MCFM & Ref. \cite{frixwz}
            & \MCFM & Ref. \cite{frixwz} & \MCFM & Ref. \cite{frixzz} \\
\hline
\hline
Born [pb]& $64.10$ & $64.11$ & $14.64$ & $14.61$ & $10.26$ & $10.25$ & $9.76$ & $9.75$\\
\hline
Full [pb]& $99.03$ & $99.03$ & $27.18$ & $27.11$ & $18.95$ & $18.91$ & $13.3$ & $13.2$\\
\hline
\end{tabular}
\end{center}
\caption{Total cross section for the various di-boson processes
at $\sqrt{s}=16$~TeV.
For the cases of $W^\pm Z$ and $W^+W^-$ production
the electroweak parameters are,
$m_W = 80.0~{\rm GeV}$, $m_Z = 91.17~{\rm GeV}$,
$\alpha_{\rm em}^{-1}=128$ and 
$\cos\theta_w = {m_W}/{m_Z}$\cite{frixww,frixwz}.
For $Z$ pairs the input is instead,
$m_Z = 91.18~{\rm GeV}$, $\sin^2\theta_w = 0.228$ and
$\alpha_{\rm em}^{-1}=128$\cite{frixzz}.}
\label{comp16}
\end{table}

\begin{table}
\renewcommand\arraystretch{1.1}
\begin{center}
\begin{tabular}{|c||c|c||c|c||c|c||c|c|}
\hline
$\sqrt{s}=40$~TeV & \multicolumn{2}{c||}{$W^+W^-$} & \multicolumn{2}{c||}{$ZW^+$} 
& \multicolumn{2}{c||}{$ZW^-$} & \multicolumn{2}{c|}{$ZZ$} \\
\cline{2-9}
($pp$)& \MCFM & Ref. \cite{frixww} & \MCFM & Ref. \cite{frixwz}
            & \MCFM & Ref. \cite{frixwz} & \MCFM & Ref. \cite{frixzz} \\
\hline
\hline
Born [pb] & $148.9$ & $149.0$ & $33.56$ & $33.49$ & $25.58$ & $25.54$ & $23.5$ & $23.5$ \\
\hline
Full [pb]& $254.1$ & $254.1$ & $71.86$ & $71.71$ & $54.78$ & $54.67$ & $34.1$ & $33.9$ \\
\hline
\end{tabular}
\end{center}
\caption{Total cross section for the various di-boson processes
at $\sqrt{s}=40$~TeV. Input parameters are given in the caption to
Table~\ref{comp16}.}
\label{comp40}
\end{table}

\subsection{Results for Tevatron Run II and the LHC}

In order to update the currently available predictions, we use 
modern values of the particle masses and widths, as given in~\cite{pdg98},
\begin{eqnarray}
&&M_W=80.41~{\rm GeV}, \qquad   \Gamma_W = 2.06~{\rm GeV}, \nonumber \\
&&M_Z=91.187~{\rm GeV}, \qquad \Gamma_Z = 2.49~{\rm GeV},
\end{eqnarray}
together with $\alpha_{\rm em}^{-1}=128.89$
and present the results using the most recent distributions of two popular sets
of structure functions, MRS98~\cite{mrs98} and
CTEQ5~\cite{cteq5}. 

The total cross-sections expected at the Tevatron and the LHC are shown in
Tables~\ref{tevtot} and~\ref{lhctot}. We have used a central gluon and
$\alpha_S(M_Z)=0.1175$ for the MRS98 parton distribution set (ft08a), 
whilst CTEQ5M has $\alpha_S(M_Z)=0.118$. 
As before, the factorization and renormalization 
scale $\mu$ is set equal to the average of the produced vector boson masses.
Note that because of changes in the structure functions and $\alpha_s$
the modern cross sections at $\sqrt{s}=14$~TeV 
lie well above the old values at $\sqrt{s}=16$~TeV given in 
Table~\ref{comp16}.

\begin{table}
\renewcommand\arraystretch{1.1}
\begin{center}
\begin{tabular}{|c||c|c||c|c||c|c|}
\hline
{\small $\sqrt{s}=2$~TeV} & \multicolumn{2}{c||}{$W^+W^-$}
& \multicolumn{2}{c||}{$ZW^+$ or $ZW^-$} & \multicolumn{2}{c|}{$ZZ$} \\
\cline{2-7}
{\small($p{\bar p}$)}& {\small MRS98} & {\small CTEQ5}
                     & {\small MRS98} & {\small CTEQ5} & {\small MRS98} & {\small CTEQ5} \\
\hline
\hline
Born [pb] & $10.0$ & $10.3$ & $1.46$ & $1.49$ & $1.22$ & $1.25$ \\
\hline
Full [pb] & $13.0$ & $13.5$ & $1.95$ & $2.01$ & $1.56$ & $1.60$ \\
\hline
\end{tabular}
\end{center}
\caption{Total cross section for the various di-boson processes
for the Tevatron Run II. Input parameters are given in the text.}
\label{tevtot}
\end{table}

\begin{table}
\renewcommand\arraystretch{1.1}
\begin{center}
\begin{tabular}{|c||c|c||c|c||c|c||c|c|}
\hline
{\small $\sqrt{s}=14$~TeV} & \multicolumn{2}{c||}{$W^+W^-$} & \multicolumn{2}{c||}{$ZW^+$}
& \multicolumn{2}{c||}{$Z W^-$} & \multicolumn{2}{c|}{$ZZ$} \\
\cline{2-9}
{\small($pp$)}& {\small MRS98} & {\small CTEQ5} & {\small MRS98} & {\small CTEQ5}
              & {\small MRS98} & {\small CTEQ5} & {\small MRS98} & {\small CTEQ5} \\
\hline
\hline
Born [pb] & $81.8$ & $86.7$ & $18.6$ & $19.9$ & $11.7$ & $12.5$ & $12.2$ & $12.9$ \\
\hline
Full [pb] & $120.6$ & $127.8$ & $31.9$ & $34.0$ & $20.2$ & $21.4$ & $16.3$ & $17.2$ \\
\hline
\end{tabular}
\end{center}
\caption{Total cross section for the various di-boson processes
for the LHC. Input parameters are given in the text.}
\label{lhctot}
\end{table}
One can see that in Run II at the Tevatron, the K-factor (the ratio of the full
next-to-leading order result to the Born level prediction) is approximately
$1.3$ for each case, whilst at the LHC it varies between $1.3$ for $Z$-pairs
and $1.7$ for $ZW^-$. The differences between the two choices of parton 
distributions considered in this paper
are of the order of 3\% 
in Run II, but about  6\% 
at the LHC.

\begin{figure}
\begin{center}
\epsfig{file=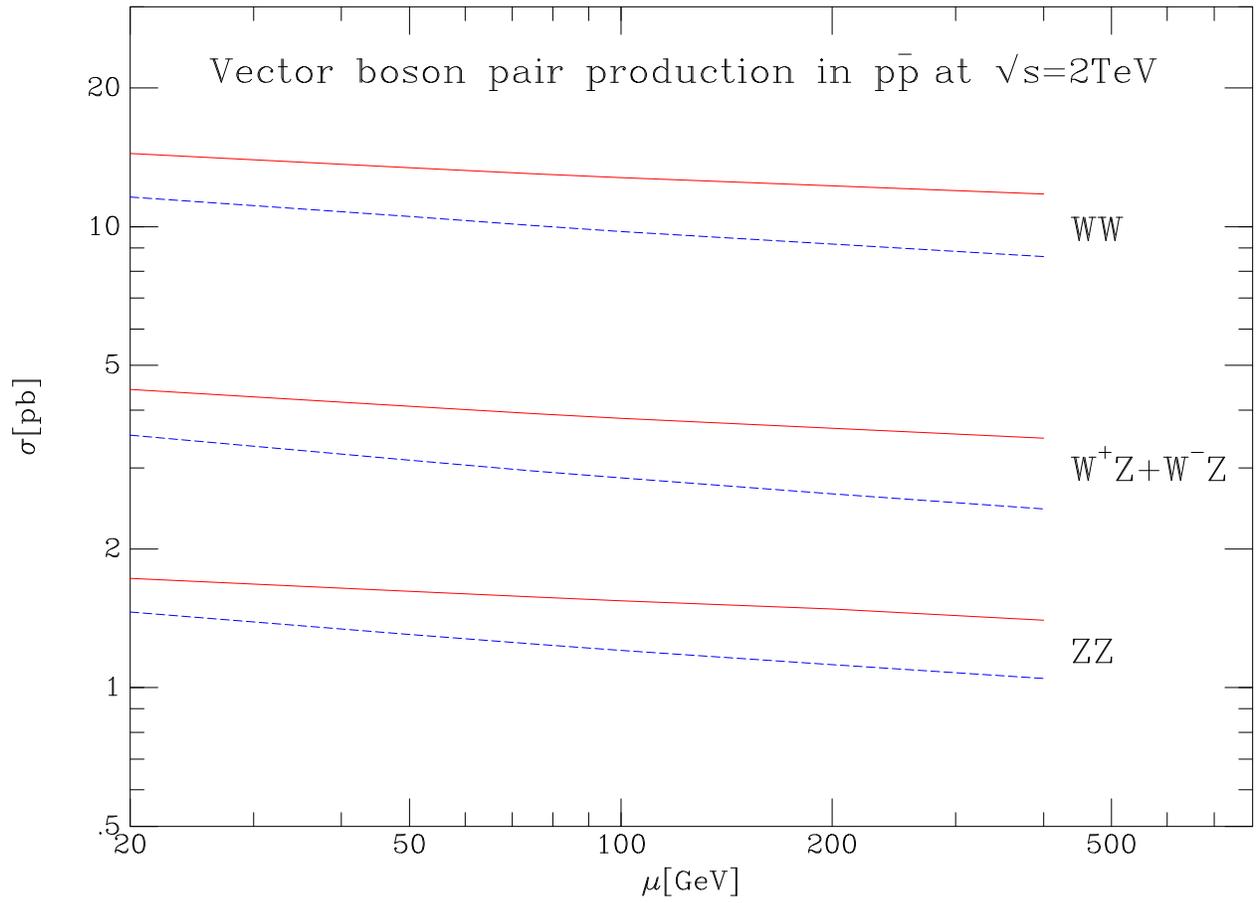,angle=270,width=16.5cm}
\end{center}
\caption{Scale dependence of vector boson pair production cross sections.}
\label{ww}
\end{figure}

\begin{figure}
\begin{center}
\epsfig{file=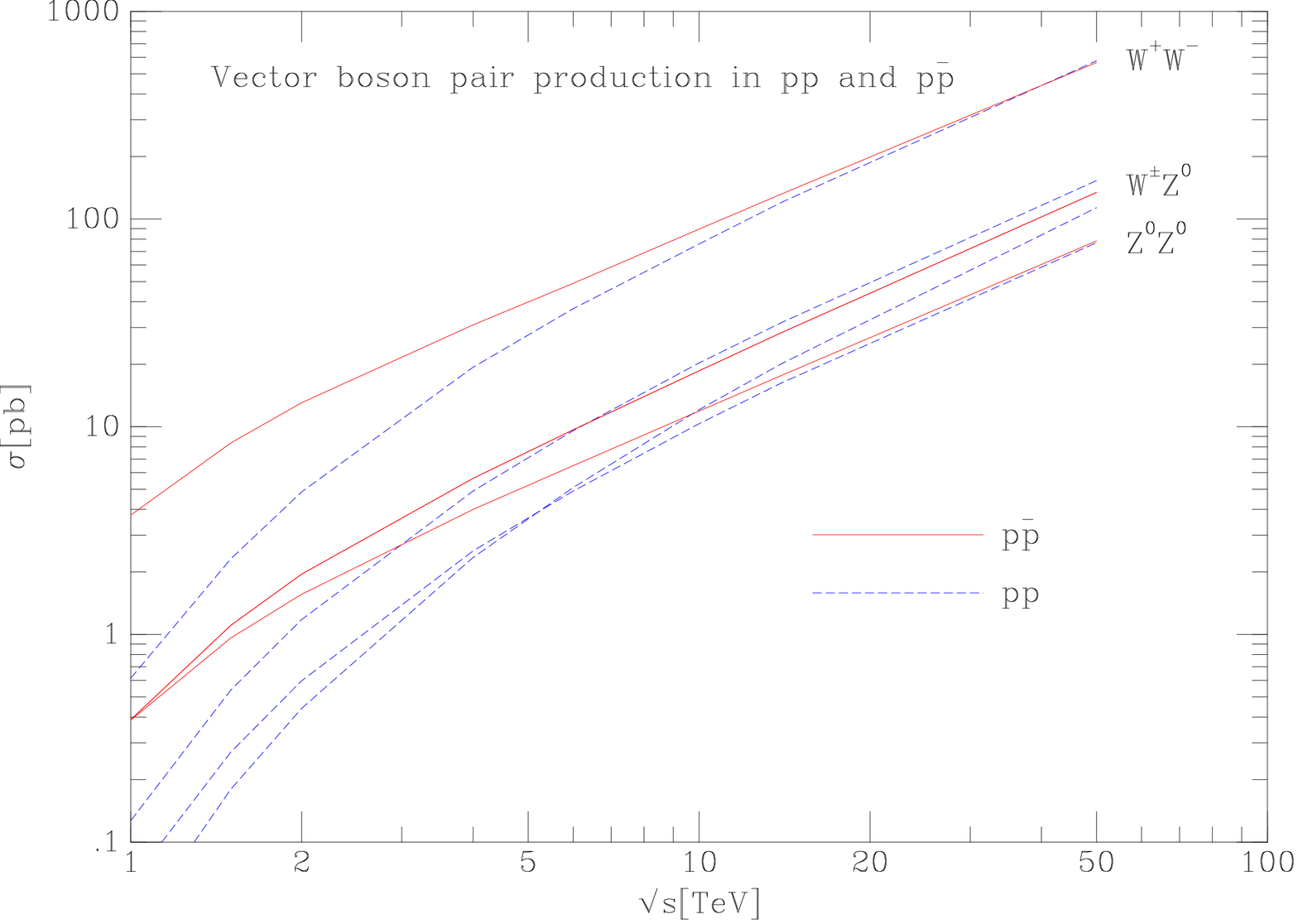,angle=270,width=16.5cm}
\end{center}
\caption{Energy dependence of vector boson pair production cross sections.
The scale $\mu$ is taken to be the average vector boson mass.}
\label{wwroots}
\end{figure}

Fig.~\ref{ww} shows the scale dependence of the cross section at 
$\sqrt{s}=2$~TeV both in leading and next-to-leading order using the MRS98
distribution. 
The growth of the cross sections with energy is shown in Fig.~\ref{wwroots},
emphasizing that at high energy vector boson pair production 
is dominated by production off sea partons. Note however 
that it is still true that $\sigma(W^+Z)>\sigma(W^-Z)$ at the energy 
of the LHC.

\section{Beyond the zero-width approximation}
\label{beyondzw}

Part of the reason for re-evaluating the vector boson pair production
cross sections is to estimate their importance as backgrounds for new physics
processes. In this context the tails of the Breit-Wigner distributions may be 
important. We are therefore motivated to go beyond the zero width 
approximation. We consider all standard model contributions
to four lepton production, rather than just those 
proceeding through the production of a pair of vector bosons. 

In the zero width approximation $q^2 = M_V^2$, the doubly-resonant 
diagrams form a gauge invariant set. If we wish to move
beyond the zero-width approximation, so that we have $q^2 \neq M_V^2$, 
gauge invariance requires that we include all diagrams which 
contribute to a given final state. This problem has been extensively 
studied in the $e^+e^-$ environment where
similar diagrams contribute to $W$-pair production~\cite{lep2}. 

\begin{figure}
\begin{center}
\begin{minipage}{12cm}
\epsfig{file=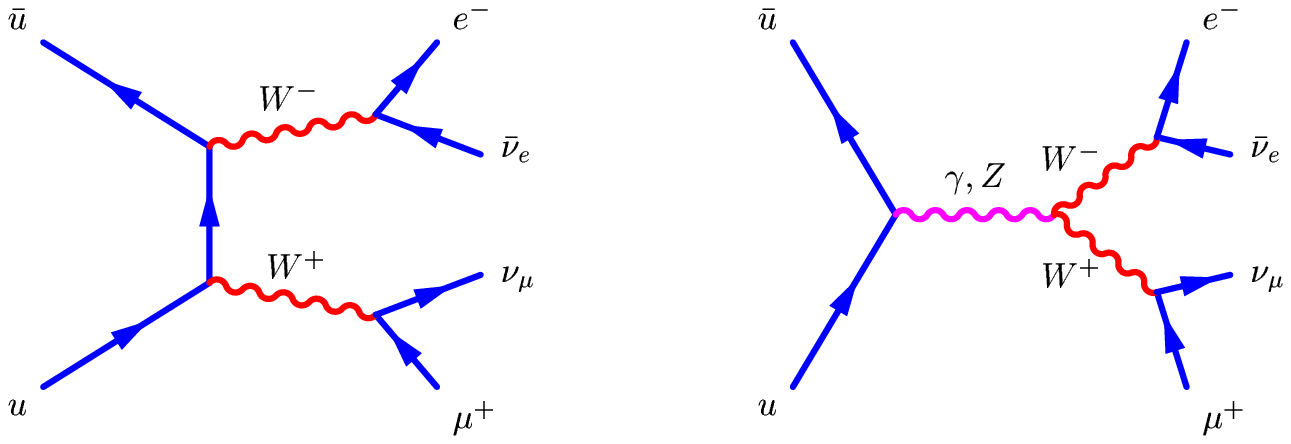,width=12cm}
\begin{center}(a)\end{center}
\vspace*{0.5cm}
\end{minipage}
\begin{minipage}{12cm}
\epsfig{file=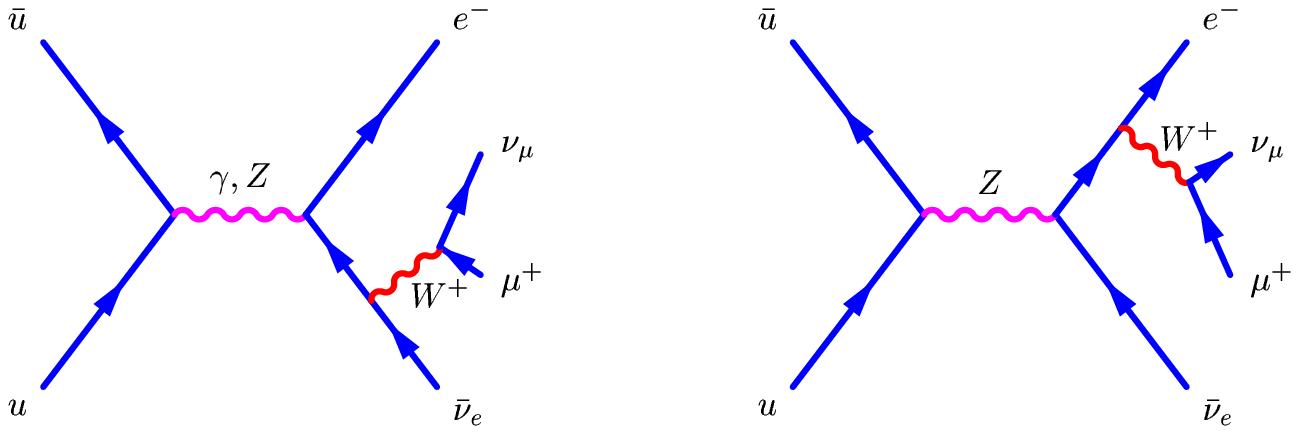,width=12cm}
\begin{center}(b)\end{center}
\end{minipage}
\end{center}
\caption{Doubly (a) and singly (b) resonant diagrams contributing to the
parton-level process $q{\bar q} \to W^+W^- \to e^- {\bar \nu_e} \nu_\mu \mu^+$.}
\label{wwdiags}
\end{figure}

In practice this means that in
addition to the diagrams containing two resonant propagators, 
we must calculate diagrams containing only a single
resonant propagator. For the case of $W$-pair production, examples of such
doubly and singly resonant diagrams are shown in Figure~\ref{wwdiags}.
To illustrate the gauge-dependence of the individual sets of diagrams one may
work in an axial gauge, taking care to include both the mixed propagators and
additional vertices that arise in these gauges. 

Even when we include all the diagrams, a second problem arises when we
introduce a width for the propagators to avoid on-shell poles. With the
modification,
\begin{displaymath}
\frac{1}{q^2-M^2} \quad \longrightarrow \quad \frac{1}{q^2-M^2+iM\Gamma},
\end{displaymath}
for each of the propagators, the amplitude is no longer gauge invariant
because we now have a mix of singly and doubly resonant diagrams.
The Breit-Wigner form of the propagator sums 
self-energy diagrams which are not separately gauge invariant.
Since the resummation of all diagrams which contribute to a given process
is not practical, several models have been proposed which allow the 
introduction of a finite width but preserve gauge invariance.

For a review of the models, the `pole-scheme'~\cite{pscheme},
the `fermion-loop scheme'~\cite{flscheme} and the
`pinch technique'~\cite{pinch}, see for example Ref.~\cite{lep2}.
Here we will adopt the simple
prescription whereby we use $1/(q^2-M^2)$ 
for each propagator initially and then
multiply the whole amplitude by,
\begin{displaymath}
\prod_{\rm props} \, \left( \frac{q^2-M^2}{q^2-M^2+iM\Gamma} \right),
\end{displaymath}
which clearly maintains gauge invariance~\cite{fudge}.
This is the correct treatment
for the doubly-resonant piece, but mistreats the singly-resonant
diagrams, primarily in the region $q^2 \sim M^2$ where the doubly resonant
diagrams dominate. This is the method that
we will use to produce our Monte Carlo results in the remainder of this paper.
Since the introduction of the 
width represents an
all-orders resummation of an partial set of diagrams, there is no
unique way to include it at a given order.

\subsection{Cross-sections to leptonic final states}

We now present our results for the various leptonic final states 
using the prescription given above.
In practice, the observed cross sections are limited by the
acceptance of the detectors in rapidity 
and in transverse momentum. We quote cross-sections 
for particular channels with cuts appropriate for Run II and the LHC. 
Specifically,
we apply cuts on the transverse momentum and rapidity of each lepton,
\setlength{\jot}{8pt}
\begin{eqnarray}
p_T > 20~{\rm GeV} &,& \qquad |\eta| < \eta_{\rm max}, \nonumber \\
\eta_{\rm max} = & \biggl\{ & {2 \qquad {\rm Run~II} \atop 2.5 \qquad {\rm LHC}~~}
\label{cuts1}
\end{eqnarray}
and a cut on the total missing transverse energy,
\begin{equation}
E_T\hspace{-13pt}\slash~~ > 25~{\rm GeV}.
\label{cuts2}
\end{equation}
\setlength{\jot}{3pt}
This suffices for $WW$ production; for the other cases we perform 
a mass cut $75< M_{l^+l^-} < 105~{\rm GeV}$.
For a more detailed study, one might tailor the cuts individually for each
process and include detector resolution effects.

Using the same input as the previous section, but now with the inclusion
of the singly-resonant diagrams, our full next-to-leading order
results with these cuts are summarized in Tables~\ref{tevcuts}
and~\ref{lhccuts}.

\begin{table}
\renewcommand\arraystretch{1.1}
\begin{center}
\begin{tabular}{|c||c|c||c|c||c|c||c|c|}
\hline
Run II & \multicolumn{2}{c||}{\small $W^-W^+ \to e^-{\bar\nu} e^+\nu$}
& \multicolumn{2}{c||}{\small $Z W^\pm  \to e^-e^+
\bar{\nu} e^\pm$}
& \multicolumn{2}{c|}{\small $ZZ \to e^-e^+ \mu^-\mu^+$} 
& \multicolumn{2}{c|}{\small $ZZ \to e^-e^+ (\nu \bar{\nu} \times 3)$} \\
\cline{2-9}
& {\small MRS98} & {\small CTEQ5} & {\small MRS98} & {\small CTEQ5} 
& {\small MRS98} & {\small CTEQ5} & {\small MRS98} & {\small CTEQ5} \\
\hline
\hline
$\sigma_{\rm cuts}^{\rm NLO}$~[fb]
 & $70.9$ & $73.5$ & $2.89$ & $2.99$ & $1.64$ & $1.70$ & $10.7$ & $11.0$\\
\hline
\end{tabular}
\end{center}
\caption{Cross sections (in fb) for various channels at the Tevatron Run II
with the cuts of~(\ref{cuts1}) and~(\ref{cuts2}).}
\label{tevcuts}
\end{table}
\begin{table}
\renewcommand\arraystretch{1.1}
\begin{center}
\begin{tabular}{|c||c|c||c|c||c|c||c|c|}
\hline
~LHC~ & \multicolumn{2}{c||}{\small $W^-W^+ \to e^-{\bar\nu} e^+\nu$}
& \multicolumn{2}{c||}{\small $ZW^+ \to e^-e^+\nu e^+$}
& \multicolumn{2}{c||}{\small $ZW^- \to e^-e^+{\bar \nu} e^-$}
& \multicolumn{2}{c|}{\small $ZZ \to e^-e^+ \mu^-\mu^+$} \\
\cline{2-9}
& {\small MRS98} & {\small CTEQ5} & {\small MRS98} & {\small CTEQ5}
& {\small MRS98} & {\small CTEQ5} & {\small MRS98} & {\small CTEQ5} \\
\hline
\hline
$\sigma_{\rm cuts}^{\rm NLO}$~[fb]
 & $514$ & $549$ & $34.0$ & $36.2$ & $22.9$ & $24.6$ & $12.0$ & $12.7$ \\
\hline
\end{tabular}
\end{center}
\caption{Cross sections (in fb) for various channels at the LHC
with the cuts of~(\ref{cuts1}) and~(\ref{cuts2}).}
\label{lhccuts}
\end{table}

\section{Examples}
In this section we present two examples to demonstrate 
the use of the $\MCFM$ Monte Carlo.

\subsection{Tri-lepton production at Run II}
One of the `gold-plated' supersymmetry discovery modes at Run II is gaugino
pair-production resulting in a tri-lepton signal, 
\begin{equation}
p \bar{p} \rightarrow \chi_1^-(\rightarrow l^{\prime -} \nu \chi_1^0) 
\, \chi^0_2(\rightarrow l^+ l^- \chi_1^0),
\end{equation}
where $\chi_1^0$ is the lightest supersymmetric particle.
In order to obtain a clean
signal, it is imperative to have a good understanding of the Standard Model
background, which is predominantly from leptonic decay of a $WZ$ pair.
There have been many previous studies of this background
(see \cite{matchev} and references therein),
primarily performed using the ISAJET~\cite{isajet} and PYTHIA~\cite{pythia}
event generators. Whilst event generators can have 
advantages over a fixed order calculation, as presently written 
ISAJET and PYTHIA do not include the $\gamma^*$ contribution
or the interference between the photon and the $Z$ in $WZ/\gamma^*$ production.

In the recent paper~\cite{matchev}, a set of cuts was proposed to further
isolate the trilepton signal. However, this study was 
based on a PYTHIA analysis which includes only the $WZ$ process
and not $W\gamma^*$ process in assessing the standard model 
four lepton background.

\begin{figure}
\begin{center}
\epsfig{file=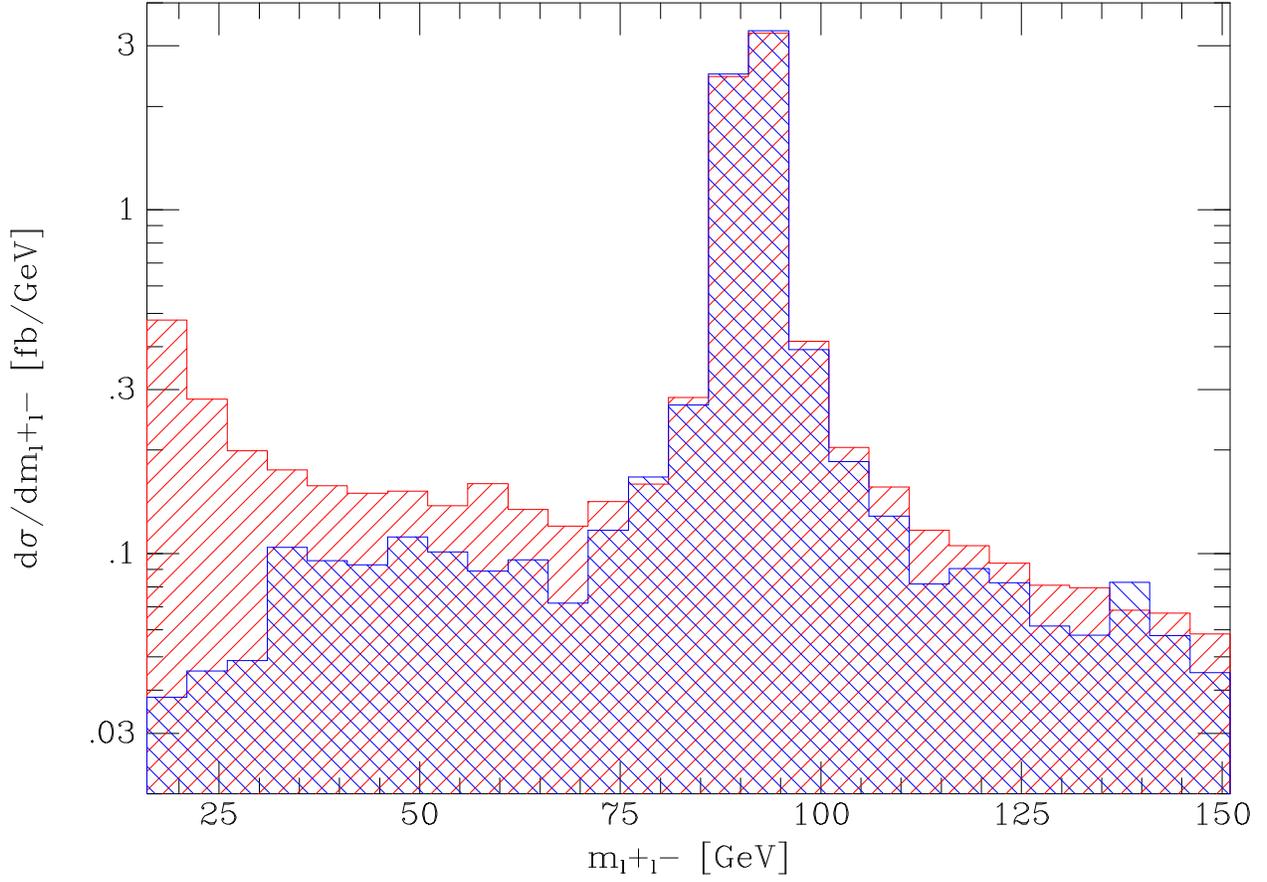,angle=270,width=16.5cm}
\end{center}
\caption{The lepton invariant mass with 
the full $Z/\gamma^*$ interference (single hatched) and with the $Z$ only 
(double hatched). Each generated event is binned once for each 
opposite-sign, same-flavour lepton pair. 
Events which give rise to two entries are binned with half the event weight.
The total cross-sections in the plot are $49.4$~fb and $42.6$~fb respectively.}
\label{wzplot}
\end{figure}

Following~\cite{matchev}, we apply the cuts,
\begin{eqnarray}
{\rm central~lepton:}&& p_T>11~{\rm GeV}, \qquad |\eta|<1, \nonumber \\
{\rm remaining~leptons:}&& p_T>7~{\rm GeV~and~}  p_T>5~{\rm GeV}, \\
E_T\hspace{-13pt}\slash~~ &>& 25~{\rm GeV}, \nonumber
\end{eqnarray}
and examine the invariant mass distribution of opposite-sign, same-flavour
lepton pairs. This differential cross-section (using the MRS98 structure
functions at $\sqrt{s}=2$~TeV) is shown in Figure~\ref{wzplot}.

We see that as the invariant mass of deviates from the $Z$-peak,
the contribution from the off-shell photon dominates. For the case presented 
in ref.~\cite{matchev} we have $m_\chi \approx 122$~GeV which gives a
signal region $10 < M_{l^+l^-} < 60$~GeV. Studies using PYTHIA 
therefore underestimate the standard model background.

\subsection{Higgs production via gluon-gluon fusion}

At the Tevatron the dominant production mode for a (Standard Model or similar)
Higgs boson is via gluon fusion, $gg \to H$ via heavy quark loops. A natural
decay mode for $140 <M_H < 180 $~GeV
is then $H \to W^{(*)} W^{(*)} \to $~leptons/jets, which has been extensively
discussed in the literature~\cite{higgs,htz}.
The most recent of these studies~\cite{htz}, performed using PYTHIA, optimized
a set of cuts to suppress the SM backgrounds for the di-lepton 
plus missing energy channel and for the like-sign lepton plus jets channel.

Here we perform a parton-level analysis for the di-lepton + missing energy 
signal. The signal is calculated using the heavy-top effective
$ggH$ vertex~\cite{dawson} and we have applied cuts (10)-(16) of~\cite{htz}
which we now describe. In addition to a standard set of cuts,
\begin{eqnarray}
\nonumber
&&p^{}_T(e)  > 10\ {\gev},\ \ |\eta^{}_e| < 1.5,\nonumber\\
&&p^{}_T(\mu_1)>10\ {\gev},\ \  p^{}_T(\mu_2)>5\ {\gev},\ \ 
|\eta^{}_\mu| < 1.5,\nonumber \\
&&m(\ell\ell)>10\ {\gev},\ \ \Delta R(\ell j)>0.4,\ \ 
\etmiss > 10\ {\gev},
\label{basic}
\end{eqnarray}
there are further cuts to reduce the various background processes. First we
apply,
\begin{equation}
\phi(\ell\ell) < 160^\circ,\ \ 
\theta(\ell\ell) < 160^\circ.
\label{phi}
\end{equation}
where $\phi(\ell\ell)$ is the azimuthal angle in the transverse 
plane and $\theta(\ell\ell)$ the three-dimensional opening-angle 
between the two leptons. We also impose,
\begin{equation}
p_T^{}(\ell\ell) > 20\ {\gev},\ \ 
\cos\theta_{\ell\ell-\etmiss} < 0.5,\ \ 
M_T^{}(\ell \etmiss) >20\ {\gev},
\label{ptll}
\end{equation}
where 
$\theta_{\ell\ell-\etmiss}$ is the relative angle between the lepton 
pair transverse momentum and the missing transverse momentum
and the two-body transverse-mass is defined for each
lepton and the missing energy as,
\begin{equation}
M_T^2(\ell \etmiss)=2 p_T(\ell)\etmiss(1-\cos\theta_{\ell-\etmiss}).
\label{tm}
\end{equation}
Further di-lepton mass cuts are,
\begin{eqnarray}
\nonumber
m(\ell\ell)&<& 78\ {\gev}\ \ {\rm for}\ e^+e^-,\mu^+\mu^-,\\ 
m(\ell\ell)&<& 110\ {\gev}\ \ {\rm for}\ e\mu,
\label{zout}
\end{eqnarray}
and we also cut on the Dittmar-Dreiner angle
$\theta^*_{\ell_1}$ for each lepton $\ell_1$,
\begin{equation}
-0.3 < \cos\theta^*_{\ell_1} < 0.8.
\label{theta}
\end{equation}
Finally, we introduce a jet veto,
\begin{eqnarray}
{\rm veto\ if}\ \ p_T^{j_1}>95\ {\gev},\ \  |\eta^{}_j| < 3,\nonumber\\ 
{\rm veto\ if}\ \ p_T^{j_2}>50\ {\gev},\ \  |\eta^{}_j| < 3,
\label{jetveto}
\end{eqnarray}
and reject events where either jet is $b$-tagged with an efficiency,
\begin{equation}
\epsilon_b = 1.1\times 57\% \  
{\rm tanh}({\eta_b\over 36.05}).
\label{btag}
\end{equation}

The results of our analysis ($\sigma_{\rm cuts}$)
at $\sqrt{s}=2$~TeV for $140 < m_H < 190$ are
shown in Table~\ref{higgstab}, where we have
employed the structure function set MRS98. Also shown
in Table~\ref{higgstab} are the total cross sections for the various channels
which serve as normalizations of our results.
The primary background in this channel is from $W^+W^-$ with smaller 
backgrounds from top pair production\footnote{This process
is implemented only at leading order in $\MCFM$.},
$WZ$ and $ZZ$ production. 
For the signal,
we choose the renormalization scale $\mu=m_H$; for the di-boson backgrounds
we again use the mean boson mass. In the case of the $t{\bar t}$
background we set $\mu=100$~GeV; order $\alpha_s^3$ corrections to 
the total top pair production cross-section are small at this scale~\cite{ESW}.

Our analysis confirms that the $WW$ process is the principal background
for Higgs production in the region $140 < m_H < 190$. Note that we have not 
included the decay $W(\rightarrow \tau \nu \rightarrow l+X)$ in 
our calculations. The comparison with ref.~\cite{htz} 
(where these effects are included) is therefore not exact. 
The effects of $\tau$ decays are stated to be small in ref.~\cite{htz}.

We find that all the backgrounds due to the di-boson processes are larger
than in ref.~\cite{htz}, primarily because we have normalized to 
the $O(\alpha_s)$ cross-sections which are about $30\%$ bigger than the Born
cross-sections, (c.f. Table~\ref{tevtot}). Note however that our signal 
cross sections are also larger than Ref.~\cite{htz} by about $20\%$,
so the net effect on $S/\sqrt{B}$ may be small. 
In addition, we find that the background from the $WZ$-class of events 
is about twice as big as ref.~\cite{htz} because of our inclusion of the
$W \gamma^*$ contribution, which was left out in ref.~\cite{htz}.
Our estimate of the $t \bar{t}$
background is smaller than in Ref.~\cite{htz}, perhaps because we 
do not include leptons from $b$-decay.  
\begin{table}
\renewcommand\arraystretch{1.1}
\begin{center}
\begin{tabular}{|c||c|c|c|c|c|c|c|c|c|c|c|c|}
\hline
Signal ($m_H$) & \multicolumn{2}{c|}{140} &
 \multicolumn{2}{c|}{150} &
 \multicolumn{2}{c|}{160} &
 \multicolumn{2}{c|}{170} &
 \multicolumn{2}{c|}{180} &
 \multicolumn{2}{c|}{190} \\
\hline
$\sigma_{\rm cuts}$ (fb) & \multicolumn{2}{c|}{$4.36$}
 & \multicolumn{2}{c|}{$5.32$} & \multicolumn{2}{c|}{$6.12$}
 & \multicolumn{2}{c|}{$5.15$} & \multicolumn{2}{c|}{$3.90$}
 & \multicolumn{2}{c|}{$2.47$} \\
$\sigma_{\rm total}$ (pb) & \multicolumn{2}{c|}{$0.181$}
 & \multicolumn{2}{c|}{$0.206$} & \multicolumn{2}{c|}{$0.215$}
 & \multicolumn{2}{c|}{$0.180$} & \multicolumn{2}{c|}{$0.143$}
 & \multicolumn{2}{c|}{$0.0976$}\\
\hline
\hline
Background & \multicolumn{3}{c|}{~~~$WW$~~~} & \multicolumn{3}{c|}{$t{\bar t}$}
 & \multicolumn{3}{c|}{~~~~$ZZ$~~~~}& \multicolumn{3}{c|}{$WZ$}\\
\hline
$\sigma_{\rm cuts}$ (fb) & \multicolumn{3}{c|}{$185$}
 & \multicolumn{3}{c|}{$9.55$} & \multicolumn{3}{c|}{$2.48$}
 & \multicolumn{3}{c|}{$10.4$} \\
$\sigma_{\rm total}$ (pb) & \multicolumn{3}{c|}{$13.0$}
 & \multicolumn{3}{c|}{$6.82$} & \multicolumn{3}{c|}{$1.56$} 
 & \multicolumn{3}{c|}{$3.96$}\\
\hline
\end{tabular}
\end{center}
\caption{Signal and background cross-sections for a Higgs search with
di-lepton final states. The $\tau^+\tau^-$ background is
negligible with these cuts. For comparison, the total cross sections
(without leptonic decays from the vector bosons) are also shown.}
\label{higgstab}
\end{table}

\section{Conclusions}
We have presented a Monte Carlo program for vector-boson pair production at
hadron colliders, including for the first time the complete
${\cal O}(\alpha_s)$ corrections with leptonic decay correlations.

We have employed this program to calculate the total di-boson cross-sections,
first as a cross-check with existing results in the literature and secondly in
order to provide an update of predictions for Run II at the Tevatron and 
for the LHC, including
the latest structure functions and strong coupling. The cross sections are
larger than previous estimates in the literature.

The advantage of this Monte Carlo is only realized when cuts are applied to
the final-state leptons. This is primarily of importance when estimating
Standard Model backgrounds to new physics, which is especially crucial at
the Tevatron. For this reason we have provided two examples of such uses in
Run II, tri-lepton production as a SUSY signal and a di-lepton analysis for an
intermediate-mass Higgs search.

\begin{appendix}
\section{Appendix: Amplitudes for four fermion processes}

We first introduce a separation of the total $n$-particle tree-level amplitude
into its doubly- and singly-resonant components,
\begin{equation}
\label{overall}
\cA_n^\tree = \cC^{\rm coup} \left(
 \cA_{n,{\rm DR}}^\tree + \cA_{n,{\rm SR}}^\tree \right),
\end{equation}
where $\cC^{\rm coup}$ is an overall coupling factor depending on the di-boson
process under consideration. The contributions $\cA_{n,{\rm DR}}^\tree$ were
first calculated in~\cite{gunkun} although we will
closely follow the more recent notation and approach of~\cite{dks}.
In particular, the amplitudes are presented in terms of particle momenta $p_i$
that are all outgoing, so that momentum conservation implies $\sum_i p_i = 0$.

\subsection{$W^+W^-$ final states}
For this process we label the particles as,
\begin{displaymath}
0 \to q_1 \qb_2 \, W^- \left( \to \ell_3 \nub_4 \right) 
 \, W^+ \left( \to \ellb_5 \nu_6 \right),
\end{displaymath}
where the leptons are not necessarily of the same flavour, and we write the
process in this manner to remind the reader that $q_1$ represents an outgoing
quark (so that when we cross to obtain the desired result, it becomes an
incoming anti-quark). Then from~\cite{dks} we see that the
non-vanishing doubly-resonant helicity amplitudes for up-quark annihilation are
(with labels suppressed where possible),
\begin{eqnarray}
\cA_{6,{\rm DR}}^\tree(u_1^L, \ub_2^R) &=& \left(
A_6^{\tree,a}(1,2,3,4,5,6) + C_{L,u} A_6^{\tree,b} (1,2,3,4,5,6)
\right) \cP_{34} \cP_{56}, \nonumber \\
\cA_{6,{\rm DR}}^\tree(u_1^R, \ub_2^L) &=& 
C_{R,u} A_6^{\tree,b}(2,1,3,4,5,6) \, \cP_{34} \cP_{56},
\label{wwdr}
\end{eqnarray}
with the sub-amplitudes $A_6^{\tree}$ given by equations (2.8) and (2.9)
of~\cite{dks}.
The $\cP_{ij}$ are propagator factors given by,
\begin{equation}
\cP_{ij} = \frac{s_{ij}}{s_{ij}-M_{ij}^2},
\end{equation}
with $M_{12}=M_Z$ and $M_{34}=M_{56}=M_W$.
The couplings that appear in these amplitudes are,
\begin{eqnarray}
\cC^{\rm coup}&=&\left( \frac{e^2}{\sstw} \right)^2, \\
C_{L,\{ {u\atop d} \}} &=& \pm 2 Q_{\{ {u\atop d} \}} \sstw 
+ ( 1 \mp 2 Q_{\{ {u\atop d} \}} \sstw ) \cP_{12}, \\
C_{R,\{ {u\atop d} \}} &=& \pm 2 Q_{\{ {u\atop d} \}} \sstw \left( 1 
- \cP_{12} \right),
\end{eqnarray}
where $Q_i$ is the electric charge in units of the positron charge.
For the additional singly-resonant diagrams we find,
\begin{eqnarray}
\lefteqn{\cA_{6,{\rm SR}}^\tree(u_1^L,\ub_2^R) =
    2\sstw \times \Bigl\{ } \nonumber \\
&& \left( \cP_{56}
A_6^{\tree,a}(3,4,6,5,2,1) + \cP_{34}
A_6^{\tree,a}(6,5,1,2,4,3) \right) v_{L,n} v_{L,u} \cP_{12} \nonumber \\
 &+& \left( \cP_{56}
A_6^{\tree,a}(3,4,1,2,5,6) + \cP_{34}
A_6^{\tree,a}(6,5,3,4,2,1) \right) 
\left(Q_u Q_e+v_{L,e} v_{L,u} \cP_{12} \right) \Bigr\} , \nonumber \\
\lefteqn{\cA_{6,{\rm SR}}^\tree(u_R,\ub_L) =
\cA_{6,{\rm SR}}^\tree(u_L,\ub_R) \left( L \leftrightarrow R,
 1 \leftrightarrow 2 \right),}
\end{eqnarray}
where we have introduced a further set of scaled couplings~\cite{dks},
\begin{eqnarray}
v_{L,e} = { -1 - 2 Q_e \sin^2 \theta_W \over \sin 2 \theta_W },&& \qquad
v_{R,e} = -{ 2 Q_e \sin^2 \theta_W \over  \sin 2 \theta_W }, \\  
v_{L,q} = { \pm 1 - 2 Q_q \sin^2 \theta_W \over  \sin 2 \theta_W },&& \qquad
v_{R,q} = -{ 2 Q_q \sin^2 \theta_W \over \sin 2 \theta_W }, \\
v_{L,n} = { 1 \over \sin 2 \theta_W }.
\end{eqnarray}
Note that here the sub-amplitudes needed for the singly-resonant diagrams
are exactly those introduced in~\cite{dks} to
describe the doubly-resonant diagrams. This is because the diagrams are
topologically equivalent and (modulo couplings) we can obtain the singly
resonant amplitudes by simply re-labelling the external legs. The down-quark
amplitudes may be obtained simply by symmetry,
\begin{eqnarray}
\cA_{6,{\rm DR}}^\tree(d,\db) &=&
\cA_{6,{\rm DR}}^\tree(u,\ub) \left(u \leftrightarrow d,
 3 \leftrightarrow 6, 4 \leftrightarrow 5 \right), \nonumber \\
\cA_{6,{\rm SR}}^\tree(d,\db) &=&
\cA_{6,{\rm SR}}^\tree(u,\ub) \left(u \leftrightarrow d \right).
\end{eqnarray}
As described in~\cite{dks}, the doubly-resonant amplitudes for the
process with an additional gluon radiated from the quark line are exactly
analogous to~(\ref{wwdr}) and require the introduction of functions
$\cA_{7,\tree}$. However, unlike the 6-particle amplitudes, these functions are
not sufficient to describe the singly-resonant diagrams. In this case, 
initial state gluon radiation in the singly-resonant diagrams
would correspond to final-state radiation in the equivalent doubly-resonant
diagrams and thus we need to introduce a new set of sub-amplitudes.
For a positive helicity gluon of momentum $p_7$ we find,
\begin{eqnarray}
\lefteqn{
\cA_{7,{\rm SR}}^\tree(u^L_1,\ub^R_2, g_7^+) =
 2\sstw \times \Bigl\{ } \nonumber \\
&& \left( \cP_{56}
B_7^{\tree,a}(1,2,3,4,5,6,7) + \cP_{34}
B_7^{\tree,b}(2,1,3,4,5,6,7) \right) v_{L,n} v_{L,u} \cP_{127} \nonumber \\
&+& \left( \cP_{56}
B_7^{\tree,b}(2,1,6,5,4,3,7) + \cP_{34}
B_7^{\tree,a}(1,2,6,5,4,3,7) \right)
\left(Q_u Q_e+v_{L,e} v_{L,u} \cP_{127} \right) \Bigr\}, \nonumber \\
\end{eqnarray}
where the new functions are defined by,
\begin{eqnarray}
B_7^{\tree,a} &=& 
   {i \, \spa3.6 \spab1.{(2+7)}.4 \spab1.{(3+6)}.5
 \over {\spa1.7  \spa2.7 s_{56} \; t_{127} \ t_{356}}}, 
\end{eqnarray}
and,
\begin{eqnarray}
B_7^{\tree,b} &=& 
   {-   i \, \spa2.6 \spb4.5 \left[\spa1.2 \spab3.{(5+4)}.1-\spa2.7
   \spab3.{(5+4)}.7\right]
 \over{\spa1.7 \spa2.7   s_{34}  t_{127}  t_{345}}}. 
\end{eqnarray}
Our definition of the spinor products follows ref.~\cite{dks}.
The remaining amplitudes are obtained as above, with the additional
(negative helicity) gluon corresponding, as in the doubly-resonant case,
to the operation $-{\rm flip}_1$ defined in~\cite{dks}.
As will be the case for all the amplitudes in this appendix, the loop
contributions for the singly-resonant diagrams may be simply obtained by
following the prescription~(3.20) of~\cite{dks}.

\subsection{$W^\pm Z$ final states}
Here we label the two processes slightly differently,
\begin{eqnarray}
0 &\to& q_1 \qb_2 \, W^- \left( \to \ell_3 \nub_4 \right) 
 \, Z \left( \to \ellb_5 \ell_6 \right), \nonumber \\
0 &\to& q_1 \qb_2 \, W^+ \left( \to \nu_3 \ellb_4 \right) 
 \, Z \left( \to \ellb_5 \ell_6 \right),
\end{eqnarray}
in order to simplify the form of the amplitudes and the overall coupling 
(c.f. ref.~\ref{overall}) is,
\begin{displaymath}
\cC^{\rm coup}=\left( \frac{e^2}{\sin\theta_W} \right)^2. \\
\end{displaymath}
The doubly-resonant amplitude
for a left-handed decay of the $Z$ is,
\begin{eqnarray}
\cA_{6,{\rm DR}}^\tree(q_1,\qb_2,\ell_6^L) &=& 
A_6^{\tree,a}(1,2,3,4,5,6)
 \left( v_{L,q_2} v_{L,e} \cP_{56}+Q_{q_2} Q_e \right) \cP_{34} \nonumber \\
&+& A_6^{\tree,a}(1,2,6,5,4,3)
 \left( v_{L,q_1} v_{L,e} \cP_{56}+Q_{q_1} Q_e \right) \cP_{34} \nonumber \\
&\pm& A_6^{\tree,b}(1,2,3,4,5,6) 
\left( v_{L,e} \cot\theta_W \cP_{56} + Q_e \right) \cP_{12} \cP_{34},
\end{eqnarray}
where the masses in the propagators are now $M_{12}=M_{34}=M_W$ and
$M_{56}=M_Z$. For $W^+$ production we have $q_1=d$, $q_2=u$ and
the third line has a positive contribution, whilst $W^-$ corresponds to
$q_1=u$, $q_2=d$ and a negative sign. This reduces to the form given
in~\cite{dks} if we set $Q_e = 0$ to neglect the virtual photon diagrams. 
The right-handed amplitude is obtained by a symmetry transformation,
\begin{equation}
\cA_{6,{\rm DR}}^\tree(q_1,\qb_2,\ell_6^R) = 
\cA_{6,{\rm DR}}^\tree(q_1,\qb_2,\ell_6^L)
\left( v_{L,e} \leftrightarrow v_{R,e}, 5 \leftrightarrow 6 \right). 
\end{equation}

The singly resonant diagrams are somewhat more complicated. With a $Z$
propagator we can couple both electrons and neutrinos, while $\gamma^*$ may
only couple directly to the electrons. In addition, if the final state electrons
are both left-handed, there is a contribution from a diagram containing two
$W$ propagators.
In total we obtain,
\begin{eqnarray}
\cA_{6,{\rm SR}}^\tree(q_1,\qb_2,\ell_6^L) &=& 
 \left( v_{L,1} A_6^{\tree,a}(3,4,6,5,2,1)
 + v_{L,2} A_6^{\tree,a}(3,4,1,2,5,6) \right)
 \cP_{56} \cP_{12} v_{L,e} \nonumber \\
&+& \left( c_1 A_6^{\tree,a}(3,4,6,5,2,1)
 + c_2 A_6^{\tree,a}(3,4,1,2,5,6) \right) \cP_{12} Q_e^2 \nonumber \\
&+& \left( c_1 A_6^{\tree,a}(6,5,1,2,4,3)
 + c_2 A_6^{\tree,a}(6,5,3,4,2,1) \right) \frac{\cP_{12} \cP_{34}}{2\sstw},
\end{eqnarray}
where the newly introduced couplings $v_{L,i}$ and $c_i$ depend upon the process
under consideration and are given by,
\begin{eqnarray}
W^+ &:& c_1=0, \qquad c_2=1, \qquad v_{L,1}=v_{L,n}, \qquad v_{L,2}=v_{L,e},
\nonumber \\
W^- &:& c_1=1, \qquad c_2=0, \qquad v_{L,1}=v_{L,e}, \qquad v_{L,2}=v_{L,n}.
\end{eqnarray}
The right-handed contribution is similar, but does not contain the corresponding
final term,
\begin{eqnarray}
\cA_{6,{\rm SR}}^\tree(q_1,\qb_2,\ell_6^R) &=& 
 \left( v_{L,1} A_6^{\tree,a}(3,4,5,6,2,1)
 + v_{L,2} A_6^{\tree,a}(3,4,1,2,6,5) \right)
 \cP_{56} \cP_{12} v_{R,e} \nonumber \\
&+& \left( c_1 A_6^{\tree,a}(3,4,5,6,2,1)
 + c_2 A_6^{\tree,a}(3,4,1,2,6,5) \right) \cP_{12} Q_e^2.
\end{eqnarray}
We now turn to the graphs including gluon radiation. For the doubly resonant
contribution with a positive helicity gluon we find a similar structure,
\begin{eqnarray}
\cA_{7,{\rm DR}}^\tree(q_1,\qb_2,\ell_6^L,g_7^+) &=& 
A_7^{\tree,a}(1,2,3,4,5,6,7)
 \left( v_{L,q_2} v_{L,e} \cP_{56}+Q_{q_2} Q_e \right) \cP_{34} \nonumber \\
&+& A_7^{\tree,a}(1,2,6,5,4,3,7)
 \left( v_{L,q_1} v_{L,e} \cP_{56}+Q_{q_1} Q_e \right) \cP_{34} \nonumber \\
&\pm& A_7^{\tree,b}(1,2,3,4,5,6,7) 
\left( v_{L,e} \cot\theta_W \cP_{56} + Q_e \right) \cP_{127} \cP_{34},
\end{eqnarray}
whilst the singly resonant pieces again require the new amplitudes,
\begin{eqnarray}
\cA_{7,{\rm SR}}^\tree(q_1,\qb_2,\ell_6^L,g_7^+) &=& 
 \left( v_{L,1} B_7^{\tree,a}(1,2,3,4,5,6,7)
 + v_{L,2} B_7^{\tree,b}(2,1,6,5,4,3,7) \right)
 \cP_{56} \cP_{127} v_{L,e} \nonumber \\
&+& \left( c_1 B_7^{\tree,a}(1,2,3,4,5,6,7)
 + c_2 B_7^{\tree,b}(2,1,6,5,4,3,7) \right) \cP_{127} Q_e^2 \nonumber \\
&+& \left( c_1 B_7^{\tree,b}(2,1,3,4,5,6,7)
 + c_2 B_7^{\tree,a}(1,2,6,5,4,3,7) \right) \frac{\cP_{127} \cP_{34}}{2\sstw},
\end{eqnarray}
\begin{eqnarray}
\cA_{7,{\rm SR}}^\tree(q_1,\qb_2,\ell_6^R,g_7^+) &=& 
 \left( v_{L,1} B_7^{\tree,a}(1,2,3,4,6,5,7)
 + v_{L,2} B_7^{\tree,b}(2,1,5,6,4,3,7) \right)
 \cP_{56} \cP_{127} v_{R,e} \nonumber \\
&+& \left( c_1 B_7^{\tree,a}(1,2,3,4,6,5,7)
 + c_2 B_7^{\tree,b}(2,1,5,6,4,3,7) \right) \cP_{127} Q_e^2.
\end{eqnarray}
The remaining amplitudes, with the helicity of the gluon reversed, can be
obtained by the transformation,
\begin{equation}
\cA_{7,{\rm SR}}^\tree(q_1,\qb_2,\ell_6,g_7^-) = -
\cA_{7,{\rm SR}}^\tree(q_1,\qb_2,\ell_6,g_7^+) \left(
4 \leftrightarrow 6, 3 \leftrightarrow 5,
\langle ab \rangle \leftrightarrow [ab],
B_7^{\tree,a} \leftrightarrow B_7^{\tree,b} \right).
\end{equation}
\end{appendix}

\end{document}